# TOWARD A GENERAL THEORY OF CONTROL KNOBS*

E. Bjorklund, LANSCE, Los Alamos, NM 87545, USA

*Abstract*

Experience using hardware knobs as an operator interface for controlling equipment has varied considerably at different facilities. Reactions have ranged from "indispensable" at some facilities, to "virtually useless" at others. In this paper, we will attempt to outline some basic principles for successful knob implementation. These principles derive from our own experience with control knobs at the LANSCE Accelerator Complex and our recent effort to adapt our control knobs to EPICS. Topics to be covered include what the operator needs the knob to do, what the knob needs to know about the device it is controlling, and knob scaling issues. Advanced topics such as closed-loop, PID, and aggregate knobs are left as topics for future papers.

## 1 INTRODUCTION

In a sense, this paper had its beginnings in the workshop on workstations held at the 1989 ICALEPCS in Vancouver [1]. During that workshop, the topic of control knobs became one of the polarizing issues. We were surprised to discover that some facilities found knobs to be rather useless, since we have generally had good experience with our own control knobs. Based on which facilities were reporting good knob experiences and which facilities were reporting bad experiences, and without giving the matter much further thought, we concluded (wrongly) that knobs were a good operator interface for rapidly cycling machines (30 Hz and above) and a rather poor interface for slow cycling machines (1 Hz or less). In other words, the usefulness of control knobs could be judged from the type of system you had.

We now fast forward to 1995 and our experience with integrating EPICS into the LANSCE control system [2]. As with any new system, one can expect an initial adjustment period with lots of complaints and grumbling. What was interesting, however, was how many of the "EPICS sucks" comments we received from operations could be traced back to the fact that EPICS devices did not control very well from our LANSCE knob hardware. In this case, the usefulness of the control system was being judged by how well it handled knobs! Since we intended to keep both EPICS, knobs, and our jobs, we decided to revisit the topic of what it takes to make a knob implementation "useful."

## 2 PRINCIPLES

Not surprisingly, much of what we learned about knob interfaces were general principles that could apply to almost any sort of operator interface. Many of these principles, in fact, were described at the same 1989 ICALEPCS conference [3,4]. At the end, we found that everything we'd learned about useful knob interfaces could be distilled into two general principles, which we refer to as "The Principle of Instant Gratification" and "The Principle of Total Control."

### 2.1 The Principle of Instant Gratification

The Principle of Instant Gratification states, "*The operator needs to know that something has happened as soon as the knob is turned*." The most likely result of violating this principle will be "overshoot." If feedback is not instantaneous, the operator is tempted to keep turning the knob until a change is observed. This usually results in excess knob pulses moving the device past its intended setpoint.

There are several obstacles that can interfere with the Principle of Instant Gratification. One such obstacle is the speed of the readback device. Some readback devices, such as DVMs and NMRs, are inherently slow and may take seconds to respond. Another, more subtle, obstacle might be a device (such as a stepper motor) with a maximum velocity that is slower than the knob velocity. In this case, the operator may see the readback change as soon as the knob is turned, but control pulses are still being produced faster than the device is consuming them, resulting in overshoot. An even more subtle obstacle to the Principle of Instant Gratification occurs in "notify-on-change" systems which implement notification deadbands to limit network traffic. If the operator makes a "tweak" that is below the notification deadband, no change will be observed in the readback, even though change has occurred at the device.

How "instant" does "instant gratification" have to be? Our experience has been that 4-5 Hz is generally a pretty good operator response rate. 2 Hz is just at the threshold of tolerability. Other recommendations for satisfying the Principle of Instant Gratification are:

- Whenever possible, both the setpoint and readback channels should be displayed. The operator can then verify that at least the setpoint is changing, even if the readback is slow.
- When it is known that the readback device will be slow, the readback value should be flagged as "slow" in the knob display to warn the operator.

---

* Work supported by US Department of Energy

- To avoid overshoot problems, the slew rate of the knob should be limited to the maximum slew rate of the device. It is also helpful if knob commands "preempt" rather than accumulate.
- If a "notify-on-change" system uses notification deadbands, these deadbands should be disabled for a short period of time following a knob command.

## *2.2 The Principle of Total Control*

The Principle of Total Control has two parts: "*The operator must be able to 'tweak' a device to the smallest level of precision allowed by the hardware*," and "*The operator must be able to slew a device throughout its entire range in a 'reasonable' amount of time*." We abbreviate these two parts as the "tweak" and "slam" rules.

The "tweak" part of the Principle of Total Control implies that, ideally, the knob should have direct access to the raw hardware units of the controlled device (DAC counts, stepper motor pulses, etc.) So the first obstacle to the "tweak" rule is the fact that many modern control systems do not allow command access at the hardware level, preferring to work exclusively in "physics" (or "engineering") units. For that matter, it is not always possible to even obtain hardware units from some of the "smarter" devices on the market. Lacking access to the hardware units directly, the next best thing is to know what the physics value is that corresponds to a change of one hardware unit (this, of course, assumes that the conversion is linear). Another obstacle to the "tweak" rule may come from the knob hardware itself. If the pulse-per-turn ratio is too high, the knob becomes overly sensitive and it will be difficult for the operator to issue just one pulse. Some systems overcome this problem through scaling. Others have "Single Step" buttons, which guarantee exactly one pulse per push. One final point worth mentioning is that the readback device must have the same (or greater) precision as the command device. If not, there is potential for violating the Principle of Instant Gratification.

The "slam" part of the Principle of Total Control is not hard to satisfy, provided the controlled device does not have an unusually large range. Given a typical knob slew rate, a 12-bit device can easily be moved through its entire range in a reasonable amount of time and still allow the fine control needed to satisfy the "tweak" rule. A 16-bit device, however, will probably require some form of "scaling" control. All this, of course, begs the question of what we mean by a "reasonable" amount of time. Our current working limit at LANSCE is to say no more then 80 turns to full scale, although this limit may be high.

## 3 SCALING

Most knob implementations have some form of scaling control. This typically involves two or three selectable gain levels, with each level altering the gain by either a factor of two or ten. Gains may be implemented either in software, or directly in the knob hardware. For most low-resolution devices (twelve bits and under), knob scaling is mainly an operator convenience, allowing them to "tweak" or "slam" the controlled device more efficiently. As the resolution of the controlled device increases, however, scaling becomes a necessity.

To illustrate this point, consider a simple scaling problem in which a 12-bit device is to be controlled by a knob with two gain selections and a basic slew rate of 100 pulses per turn (a number which yields a fairly acceptable level of fine control). The device's range is then $2^{12} = 4096$. If the fine gain is X1 and the coarse gain is X2, then the device can be controlled throughout its entire range in about 41 turns at fine gain and 20.5 turns at coarse gain. We consider both numbers acceptable enough to meet the "slam" rule.

Now suppose we have a 16-bit device (range = 65,536). At 100 pulses per turn, a coarse gain of X10 pushes the limit a bit, but does give us an acceptable number of turns to full scale (65.5). However, the effective pulses-per-turn (1,000) at this gain would make fine control very difficult. A sixteen-bit device, therefore, requires at least two gain settings in order to satisfy both the "tweak" and "slam" rules. Continuing on in this vein, we see that a 24-bit device requires three gain levels and a 32-bit device requires four.[1]

## *3.1 Proportional Scaling*

The increasing diversity and ranges of the controlled devices makes it attractive to consider dynamic scaling algorithms. In particular, we wish to consider "proportional" scaling algorithms that allow you to more evenly divide the device's range between the available gain levels. With this in mind, let:

*Range* = The range of the controlled device,
*N* = The number of gain settings on the knob,
*PPT$_j$* = The number of pulses per turn set by the knob hardware for gain *j*, where *j*=1 is the finest gain and *j=N* is the coarsest gain.
*SF$_j$* = The computed scale factor for gain *j*.

---

[1] We are not presently aware of any 24 or 32 bit DACs on the market. There are, however, 24 and 32 bit counters, for which one could envision control points requiring an equal amount of precision. We have also seen controllers for tunable lasers that use quantum effects to set the laser frequency with eight or more significant digits.

In order to satisfy the "slam" rule, we define $NT_N$ to be the maximum number of turns it should take to move the device through its full range at the coarsest gain. Given $NT_N$ we can compute what the scale factor for the coarsest gain needs to be:

$$SF_N = \frac{Range}{PPT_N \cdot NT_N} \quad (1)$$

To compute the scale factor for the next lower gain, we define $NT_{N-1}$ to be the number of turns at the next lower gain which will produce the same change as <u>one</u> turn at the current gain. Choosing a reasonable value for $NT_{N-1}$ allows us to compute $SF_{N-1}$ from a relationship which generalizes as:

$$SF_j \cdot PPT_j = NT_{j-1} \cdot SF_{j-1} \cdot PPT_{j-1} \quad (2)$$

Recursively applying equation (2) down to the finest gain setting ($j=1$), and substituting equation (1) for $SF_N$ in the final result, we arrive at:

$$Range = SF_1 \cdot PPT_1 \prod_{j=1}^{N} NT_j \quad (3)$$

It is interesting to note in equation (3) how all the $PPT_j$ terms, except the last, have cancelled out. Note that the product, $SF_1 PPT_1$, tells us whether or not our $NT_j$ choices will allow us to satisfy the "tweak" rule at the finest gain.

For proportional scaling, we want to set $NT_1 = NT_2 = \ldots = NT_N$ to insure that each gain setting does its fair share of work when spanning a large range. We call this uniform $NT$ the *TurnRatio*. By setting the finest scale factor ($SF_1$) to a value consistent with the "tweak" rule (typically $SF_1=1.0$), we can compute the *TurnRatio* from equation (3):

$$TurnRatio = \sqrt[N]{\frac{Range}{SF_1 \cdot PPT_1}} \quad (4)$$

Computed this way, the *TurnRatio* gives a good idea of how well you will be able to satisfy the "slam" rule. For example, note that a 32-bit device, on a 100 pulse per turn knob, with four gain settings, will have a *TurnRatio* of approximately 81 and a coarse scale factor ($SF_4$) of 530,542 (assuming no hardware scaling and the fine scale factor $SF_1=1.0$).

It is probably worth noting that very large scale factors can lead to (very large) overshoot problems, particularly if the scale factor is an "unusual" number like 530,542. This problem can be mitigated by adjusting the scale factors (probably at the expense of the "slam" rule) such that a single knob pulse scales to a power of ten (in <u>physics</u> units).

## 4 CONTROL SYSTEM ASSISTANCE

As indicated in the previous sections, the underlying control system can provide a lot of assistance (or hindrance) to a successful knob implementation. One important service the control system can provide is the ability to abort or override a command in progress. This is particularly important for command devices that "slew" or "ramp" their outputs (either via hardware or software). Another important service is providing a method for associating command devices with their relevant readback devices. This linkage could be accomplished by naming conventions, or through the control system database.

Other useful information the control system can provide includes: 1) the device range, including minimum and maximum control values, 2) the maximum slew rate for a device, 3) the device's minimum increment value, 4) units and precision for readback displays and, 5) some indication of whether a readback device is slow or fast.

## 5 CONCLUSIONS

Our recent research and past experience with control knobs at LANSCE leads us to the following conclusions: 1) Hardware control knobs can be a very useful and intuitive operator interface tool. 2) Implementing the software (and hardware) for successful knobs is not a trivial task, and 3) The features of the underlying control system can contribute significantly to the success or failure of a knob implementation.

We hope that this paper can provide a starting point for new facilities thinking about incorporating control knobs into their operator interface and a springboard for future discussion (and possibly ICALEPCS papers) on control knob issues.